# Imaging Electron Motion in a Few Layer MoS$_2$ Device

S Bhandari[1], K Wang[2], K Watanabe[3], T Taniguchi[3], P Kim[1,2] and R M Westervelt[1,2]

[1]School of Engineering and Applied Sciences, Harvard University, Cambridge, MA 02138, U.S.A.
[2]Department of Physics, Harvard University, Cambridge, MA 02138, U.S.A
[3]National Institute for Materials Science, 1-1 Naniki, Tsukuba, 305-0044, Japan

Email: sbhandar@fas.harvard.edu

**Abstract**. Ultrathin sheets of MoS$_2$ are a newly discovered 2D semiconductor that holds great promise for nanoelectronics. Understanding the pattern of current flow will be crucial for developing devices. In this talk, we present images of current flow in MoS$_2$ obtained with a Scanned Probe Microscope (SPM) cooled to 4 K. We previously used this technique to image electron trajectories in GaAs/AlGaAs heterostructures and graphene. The charged SPM tip is held just above the sample surface, creating an image charge inside the device that scatters electrons. By measuring the change in resistance $\Delta R$ while the tip is raster scanned above the sample, an image of electron flow is obtained. We present images of electron flow in an MoS$_2$ device patterned into a hall bar geometry. A three-layer MoS$_2$ sheet is encased by two hBN layers, top and bottom, and patterned into a hall-bar with multilayer graphene contacts. An SPM image shows the current flow pattern from the wide contact at the end of the device for a Hall density $n = 1.3 \times 10^{12}$ cm$^{-2}$. The SPM tip tends to block flow, increasing the resistance $R$. The pattern of flow was also imaged for a narrow side contact on the sample. At density $n = 5.4 \times 10^{11}$ cm$^{-2}$; the pattern seen in the SPM image is similar to the wide contact. The ability to image electron flow promises to be very useful for the development of ultrathin devices from new 2D materials.

## 1. Introduction

With the discovery of two-dimensional (2D) materials, Scanning Probe Microscope (SPM) imaging has become an important tool in shedding light on the physics of devices made from these devices at the nanoscale. SPM imaging technique has been used to image electron trajectories in GaAs/AlGaAs heterostructures [1-9] and graphene [10,11]. A recently discovered material is an ultrathin sheet of MoS$_2$ which forms a 2D semiconductor. MoS$_2$ displays unique electronic properties such as thickness-dependent band structure [13-15], spin-valley physics, and valley Hall effects [16-18]. It also shows a good carrier mobility, compared with other 2D semiconductors [15,18-22]. At liquid-He temperatures, the mobility in MoS$_2$ is limited, as a result of scattering by surface impurities and the substrate [21]. Little is known about the sources of scattering and disorder in MoS$_2$. It is speculated that intrinsic factors such as lattice defects and extrinsic factors such as charged impurities and substrate adsorbates contribute to the scattering [20, 23]. Atomically flat hexagonal boron nitride (hBN) substrate and top layers greatly improve the mobility in graphene [24] and MoS$_2$ [20]. However, the mobility in hBN-encapsulated MoS$_2$ is still lower than the theoretically predicted value [20]. Recent reports have suggested that interfacing MoS$_2$ with multilayer graphene greatly lowers the contact resistance [20,25].

In this paper, we use a cooled SPM to image electron flow in a three-layer $MoS_2$ sheet. The $MoS_2$ device is encapsulated by hBN layers on the top and bottom and patterned into a hall-bar with multilayer graphene contacts. Using the SPM, we observe images of current flow in $MoS_2$ cooled to 4.2 K. A charged SPM tip held above the sample surface, creates an image charge inside the $MoS_2$ device that scatters electrons. By measuring the change in resistance $\Delta R$ while the tip is raster scanned above the sample, an image of electron flow is obtained. The SPM images show an electron flow pattern that peaks at the center of a contact and decays exponentially into the device. The SPM images were taken at two different contacts – a wide contact at the end of the device and a narrow contact along the side. The electron density $n$ inside the device was measured using the Hall effect. For the wide contact at $n = 1.3 \times 10^{12}$ cm$^{-2}$, the SPM image shows a partially blocked region in the middle of the contact that exponentially decays with characteristic length is $L = 250\ nm$. Similarly, for the narrow contact at density $n = 5.4 \times 10^{11}$ cm$^{-2}$, a partially blocked region occurs with exponential decay with $L = 200\ nm$. The decay lengths are compared with the mean free path from a simple 2D electron gas model of transport.

## 2. Method

Figure 1(a) shows a schematic diagram of the set up used to image electron flow in the $MoS_2$ device with a cooled scanning probe microscope (SPM). The Hall bar is patterned from a hBN/$MoS_2$/hBN sandwich. The dimensions are 11.0x5.0 µm², with two narrow (width 1 µm) contacts along each side separated by 3.0 µm center-to-center, and wide contacts (width 3 µm) at either end. To improve the contact resistance, each of the contacts is interfaced with multilayer graphene which is then contacted by metal electrodes. The heavily doped Si substrate acts as a back-gate, covered by a 285 nm insulating layer of $SiO_2$. The back-gate capacitance is $C_G = 11.5$ nF. The density $n$ can be tuned to be in the conductive or insulating regime by applying an appropriate gate voltage $V_G$ between the backgate and the $MoS_2$. The density $n$ is obtained from measurements of the Hall resistance $R_H = 1/ne$ where e is the electron charge.

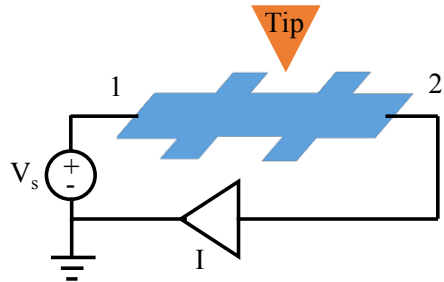

Figure 1: Experimental setup. Schematic diagram illustrating the scanning gate microscope tip above the $MoS_2$ device to image electron flow.

In our measurements, a 20 nm diameter SPM tip is scanned at a constant height above the $MoS_2$ device while the conductance of the sample is measured. As shown in Fig. 1, a dc voltage $V$ is applied between the two wide contacts while current, while the resulting current $I$ between them is measured. For this paper the electron density of the device is tuned by the backgate voltage to be in the conductive regime. An image of electron flow is obtained by displaying the change in resistance $\Delta R$ of the device as the tip is raster scanned above the sample.

## 3. Results and Discussion

SPM images of the $MoS_2$ device are shown in Fig 2(b) a.nd 2(c). Figure 1(a) shows an optical image of the hall bar device. The white outlines in Fig. 2(a) – a square and an ellipse - indicate the region of image scans. Figure 2(b) shows the resistance change $\Delta R$ as the tip is scanned over the elliptical region shown in Fig. 2(a) in the wide contact with electron density $n = 1.3 \times 10^{12}$ cm$^{-2}$. The black solid lines denote the edges of the contact. Similarly, Fig. 2(c) shows the resistance change $\Delta R$ as the tip is scanned over the square region for the narrow side contact shown in Fig. 2(a). The resistance change $\Delta R$ pattern is similar to the wide contact with resistance change peaking at the center and decaying outward. The electron density was tuned to $n = 5.4 \times 10^{11}$ cm$^{-2}$.

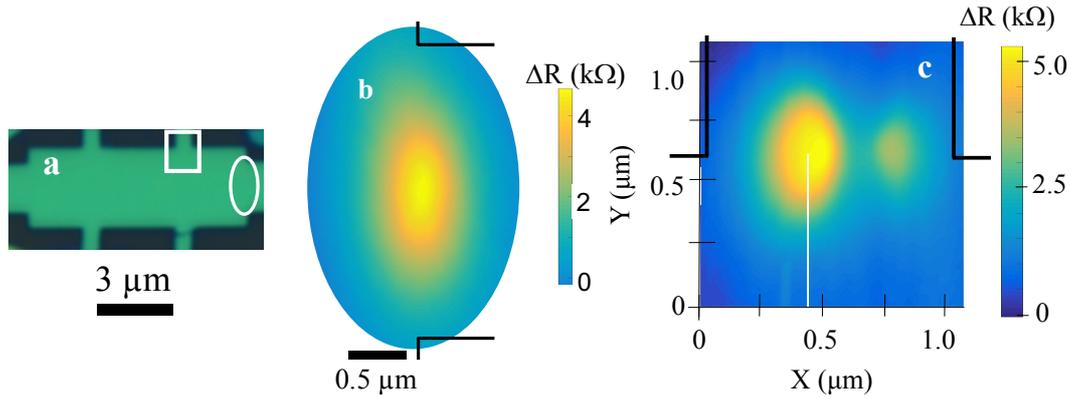

Figure 2: (a) Optical image of MoS$_2$ sample. (b) Cooled SPM image at 4.2K showing the current flow pattern through the wide contact at $n = 1.3 \times 10^{12}$ cm$^{-2}$, (c) Cooled SPM images showing current flow through the narrow contact at $n = 5.4 \times 10^{11}$ cm$^{-2}$.

A line cut of the resistance change $\Delta R$ through the middle of the partially blocked region of Fig. 2(b) is plotted in Fig. 3(a). The data is a semi-log plot of $\Delta R$ vs. the depth $X$ into the device from the contact. As shown in Fig. 3(a), $\Delta R$ drops exponentially from the end of the contact into the device with a decay length $L_1 \approx 250\ nm$. Similarly, Fig. 3(b) is a semi-log plot of $\Delta R$ vs. depth $X$ into the device for the narrow contact for Fig. 2(c). The resistance change also drops exponentially, with decay length $L_2 \approx 200\ nm$.

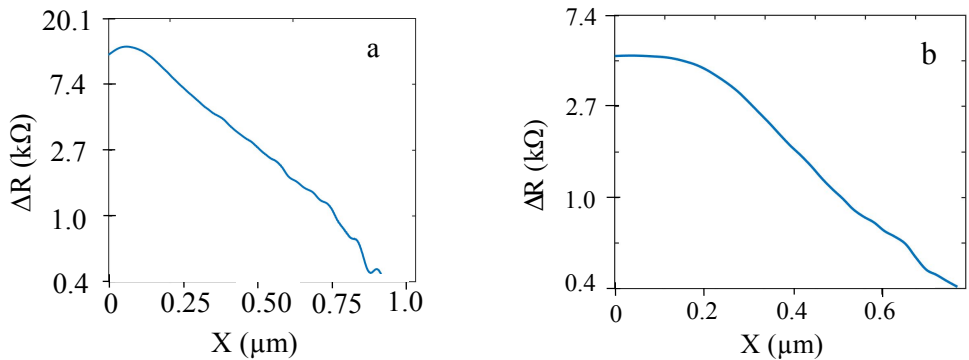

Figure 3: Semi-log plots of resistance change $\Delta R$ vs. depth $X$ into the device from the end of the contact. The plots are line cuts through the middle of the partially blocked region in Fig. 2 for (a) wide contact and (b) narrow contact. $\Delta R$ decays exponentially into the device.

To interpret these images, we compare the exponential fall off observed in the MoS$_2$ device with the mean free path estimated using a simple model. The mean free path for electrons $L_m$ is

$$L_m = v_F \tau \quad (1)$$

For a 2D free electron gas, the Fermi wavenumber $k_F$ is

$$k_F = (m^*/\hbar) v_F = (2\pi n)^{1/2} \quad (2)$$

Using Eq. 1 and 2, we find the 2D sheet resistance $\rho$

$$1/\rho = (e^2/h) k_F L_m \quad (3)$$

and the mean free path

$$L_m = \left(\frac{h}{e^2}\right) \frac{1}{\rho (2\pi n)^{1/2}} \quad (4)$$

For the wide contact the density and sheet resistance in Hall measurements are $n = 1.3 \times 10^{12}$ cm$^{-2}$ and $\rho = 2,250$ Ω. The estimated mean free path $L_m = 40$ nm is well below the measured decay length $L_1 = 250$ nm. Similarly, for the narrow contact with $n = 5.4 \times 10^{11}$ cm$^{-2}$ and $\rho = 5,130$ Ω the estimated mean free path $L_m = 27$ nm is below the measured decay length $L_1 = 200$ nm.

The estimated mean free paths are based on $n$ and $R$ measurements averaged over the whole area of the device. The discrepancy between the measured decay length and the mean free path may be due to changes in the local density $n$ and density of states near the MoS$_2$ layer edge and inside the contact. The band structure of ultrathin MoS$_2$ sheets changes with thickness [13-15] and additional states are expected at the edge. The energy bands are very sensitive to strain [26,27] which may be present in the contact regions due to the structure of the encapsulated hBN/MoS$_2$/hBN device near the contacts [28, 29]. In addition, the surface states at the edge of MoS$_2$ could change the charge density along the edge [30]. These effects could cause the mean free path to vary across the sample and differ from their average value near the contacts.

## 4. Conclusion

A cooled scanning probe microscope allows us to locally investigate the electronic motion in nanoscale structures. Recently discovered 2D materials such as MoS$_2$ offer new approaches to electronics and photonics, as well as an opportunity to probe physics in low-dimensional systems. Several unique electronic properties such as thickness/strain dependent band structure [13-15], spin-valley physics and valley Hall effects have been predicted in MoS$_2$. However, the carrier mobility has been limited to moderate values. Using our cooled SPM, we are able to locally probe electron motion at the nanoscale, inside an ultrathin MoS$_2$ device. We imaged electron motion close to two contacts, discovering a region where the contact joins the channel where the tip partially blocks electron flow. The resistance change $\Delta R$ drops exponentially with tip position as the tip moves into the sample, with a decay lengths 250 nm and 200 nm for the wide and narrow contacts. A simple model to estimate the mean free path $L_m$ using values of $n$ and $R$ for the device finds smaller values. The electronic density and band structure of MoS$_2$ are expected to vary near the edge and near the contacts, offering a possible resolution.

**Acknowledgements**
The SPM imaging research and the ray-tracing simulations were supported by the U.S. DOE Office of Basic Energy Sciences, Materials Sciences and Engineering Division, under grant DE-FG02-07ER46422. $MoS_2$ sample fabrication was supported by Air Force Office of Scientific Research contract FA9550-14-1-0268 and Army Research Office contract W911NF-14-1-0247. Growth of hexagonal boron nitride crystals was supported by the Elemental Strategy Initiative conducted by the MEXT, Japan and a Grant-in-Aid for Scientific Research on Innovative Areas No. 2506 "Science of Atomic Layers" from JSPS. Nanofabrication was performed in the Center for Nanoscale Systems (CNS) at Harvard University, a member of the National Nanotechnology Coordinated Infrastructure Network (NNCI), which is supported by the National Science Foundation under NSF award ECCS-1541959.